\documentclass[10pt,letterpaper,amsmath,amssymb]{article}
\usepackage{opex3}
\usepackage{amssymb}
\usepackage{amsmath}
\usepackage{graphicx}
\usepackage{dcolumn}
\usepackage{bm}
\DeclareFontFamily{U}{rsfs}{\skewchar\font127}
\DeclareFontShape{U}{rsfs}{m}{n}{<-6>rsfs5<6-8.5>rsfs7<8.5->rsfs10}{}
\DeclareSymbolFont{rsfs}{U}{rsfs}{m}{n}
\DeclareSymbolFontAlphabet{\mathrsfs}{rsfs}
\DeclareRobustCommand*\rsfs{\@fontswitch\relax\mathrsfs} 
\begin{document}

\title{Binary phase oscillation of two mutually coupled semiconductor lasers}

\author{Shoko Utsunomiya$^1$, Naoto Namekata$^2$, Kenta Takata$^{1,3}$, Daisuke Akamatsu$^4$, Shuichiro Inoue$^2$, and Yoshihisa Yamamoto$^{1,4,5}$}

\address{1. National Institute of Informatics, Hitotsubashi 2-1-2, Chiyoda-ku, Tokyo 101-8403, Japan 
, 2. Nihon University, 1-8-14 Kanda-Surugadai, Chiyoda-ku, Tokyo, 101-8308, Japan 3. Department of Information and Communication Engineering, The University of Tokyo, Tokyo 113-8654, Japan, 4. National Metrology Institute of Japan(NMIJ), National Institute of Advanced Industrial Science and Technology(AIST), Central 3, 1-1-1 Umezono, Tsukuba, Ibaraki 305-8563 Japan, 5. E. L. Ginzton Laboratory, Stanford University, Stanford, California 94305, USA 
}

\email{shoko@nii.ac.jp} 



\begin{abstract}
A two-site Ising model is implemented as an injection-locked laser network 
consisting of a single master laser and two mutually coupled slave lasers. 
We observed ferromagnetic and antiferromagnetic orders in the in-phase and 
out-of-phase couplings between the two slave lasers. Their phase difference  is
locked to either 0 or $\pi$ even if the coupling path is continuously 
modulated. The system automatically selects the oscillation frequency to 
satisfy the in-phase or out-of-phase coupling condition, when the mutual coupling 
dominates over the injection-locking by the master laser.  
\end{abstract}



\section{Introduction} 
\  \ 
Combinational optimization problems are ubiquitous in our modern life. Classic 
examples include protein folding in life-science and medicine, frequency assignment 
in wireless communications, navigation and routing, microprocessor circuit design, 
computer vision and voice recognition in machine learning and control in social 
network. Those combinatorial optimization problems often belong to NP (Non-deterministic Polynomial-time) hard problems, 
for which modern digital computers and future quantum computers cannot find exact
solutions in polynomial time.
\par
Three-dimensional Ising models and two-dimensional Ising models with a Zeeman 
term are also known to be NP complete/hard problems\cite{bib: Barahona 1982}.  
Quantum annealing has been proposed to solve such Ising models by utilizing 
quantum mechanical tunneling induced by a transverse field \cite{bib: Nishimori 2001}-
\cite{bib: Young 2010}. This concept was implemented with superconducting flux 
qubits in the D-Wave One / Two machines \cite{bib: Boixo 2013}. The idea of quantum 
simulation is to use a well-controlled quantum system to simulate the behavior 
of another quantum system \cite{bib: Feynman 1982}. Recent experiments using 
trapped ions \cite{bib: Friedenauer 2008}-\cite{bib: Islam 2011} have realized 
quantum simulation of Ising models.

We have proposed an alternative scheme for implementing such Ising models with 
an injection-locked laser network \cite{bib: Utsunomiya 2011, bib: Takata 2012}. 
Another physical implementation of Ising model based on degenerate optical 
parametric oscillators is also proposed \cite{bib: Wang 2013} where the binary Ising spins are represented by the 
bistable phase (0 or $\pi$) of a degenerate optical parametric oscillator (DOPO) and experimentally demonstrated \cite{bib: Marandi 2014} with $M=4$ DOPO pulses.  In principle, such injection-locked laser network and 
parametric oscillation network can be operated at room temperature and constructed 
as relatively compact systems. The Ising spins in the laser network studied in the 
present work are represented by the discrete phases ($\frac{\pi}{2}$ or $-\frac{\pi}{2}$) 
of single-mode slave lasers with respect to the master laser phase rather than 
two orthogonal polarization modes proposed in the original paper \cite{bib: Utsunomiya 2011}. 
Here, we describe the first experimental implementation of a two-site Ising 
machine using an injection-locked laser network. 
\\
\section{Ising spins in injection-locked slave lasers}
\ \ \ \ Our goal is to implement the Ising models with a Zeeman term in an injection-locked laser network. 
The Ising Hamiltonian with a Zeeman term is described as 
\begin{equation}\label{eq:1.1 Ising}
\mathrsfs{H}=\sum_{i<j}^{M}J_{ij}\sigma_{iz}\sigma_{jz}+\sum_{i}^{M} \lambda_i \sigma_{iz}
\end{equation}
In this Hamiltonian, $\sigma_{iz}$ describes an Ising spin, i.e., spin projection 
onto the z-axis. $J_{ij}$ is the interaction coefficient between spin $i$ and 
spin $j$, and $\lambda_{i}$ is a supplemental Zeeman (external field) term. 
In contrast to the standard Ising model with nearest neighbour coupling, 
here we deal with an arbitrary graph in which a vertex can be connected to any other vertices.
The injection-locked laser network proposed in our previous paper  
\cite{bib: Utsunomiya 2011, bib: Takata 2012} can find the ground state of the 
Hamiltonian, Eq. $(\ref{eq:1.1 Ising})$, through a laser phase transition. A photon 
in the lasing mode is not localized in any specific slave laser but its wavefunction 
is coherently spread over all slave lasers as partial waves of each individual photon. 
At the end of the computation, the phase configuration of such partial waves is 
expected to represent the spin configuration of a particular ground state 
$\{ \sigma_{iz}, i=1 \sim M \}$. 

\begin{figure}[tbp]
\centering\includegraphics[height=4cm,bb=12 2 610 228]{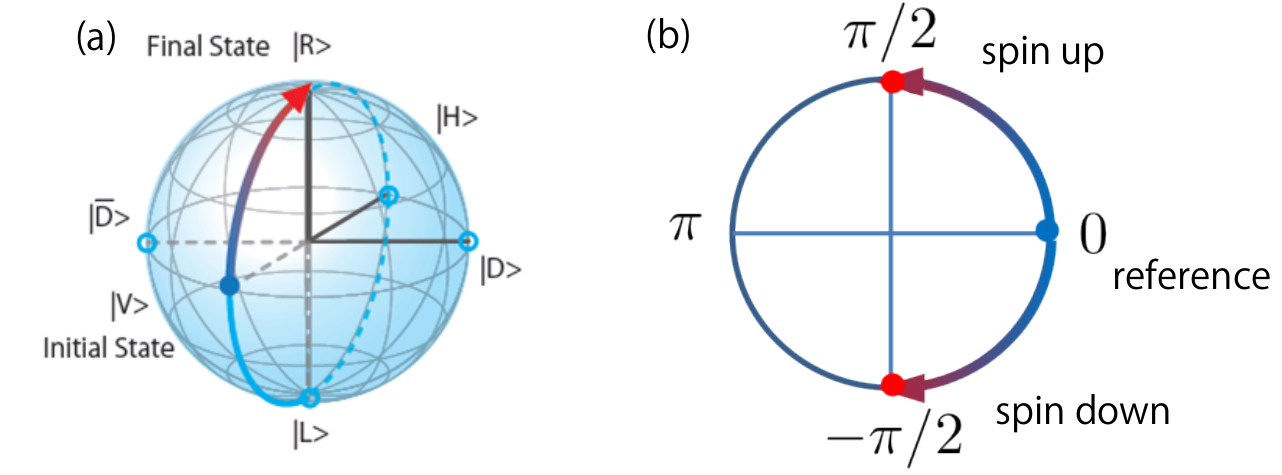}
\caption{Ising spin representation by using polarization state (a) or phase state (b) of slave lasers}\label{Fig:1}
\end{figure}
Ising spins are represented by the polarization states of the slave lasers as 
shown in Fig. \ref{Fig:1}(a), where $|R\rangle$ and $|L\rangle$ represent right 
circular (up-spin) and left circular (down-spin) polarizations, respectively. 
Each slave laser is injection locked by a master laser and prepared in the 
vertical polarized state $|V\rangle$. Then, the polarization state evolves 
towards either the right or left circular polarization after the mutual couplings 
between the slave lasers are turned on.  Such polarization evolutions are described 
as a change in the relative phase between the equal amplitude diagonal polarization 
states $|D\rangle$ and $|\bar{D}\rangle$ (see Fig. \ref{Fig:1}(a)). The 
amplitudes of two diagonal polarizations $|D\rangle$ and $|\bar{D}\rangle$ are held 
constant but their phase difference continuously evolves during the computation. 
This phase-based picture, immediately makes it apparent that we need only one 
polarization state $|D\rangle$ because the phase of the another polarization state 
$|\bar{D}\rangle$ is inversely-symmetric to the phase of $|D\rangle$ state. 
In the new model, the Ising spins $\sigma_{iz}$ are represented by the phase of 
the single slave laser mode, as shown in Fig. \ref{Fig:1}(b). The zero phase, as 
the reference, is determined by the phase of the injected master laser. When the 
master laser injection is reasonably strong and the self-oscillation frequency of 
the slave laser is identical to that of the master laser, the phase of the slave 
laser is identical to that of the master laser, i.e. $\phi_{si}=0$. The phase 
rotates to $\phi_{si}=\pm \pi/2$ from the initial phase $\phi_{si}=0$ due to the 
mutual coupling between the slave lasers , which corresponds to the Ising spin 
rotating to the up or down state.
\\
\section{Experimental setup}
\begin{figure}[htbp]
\centering\includegraphics[width=14cm, bb=0 0 701 519]{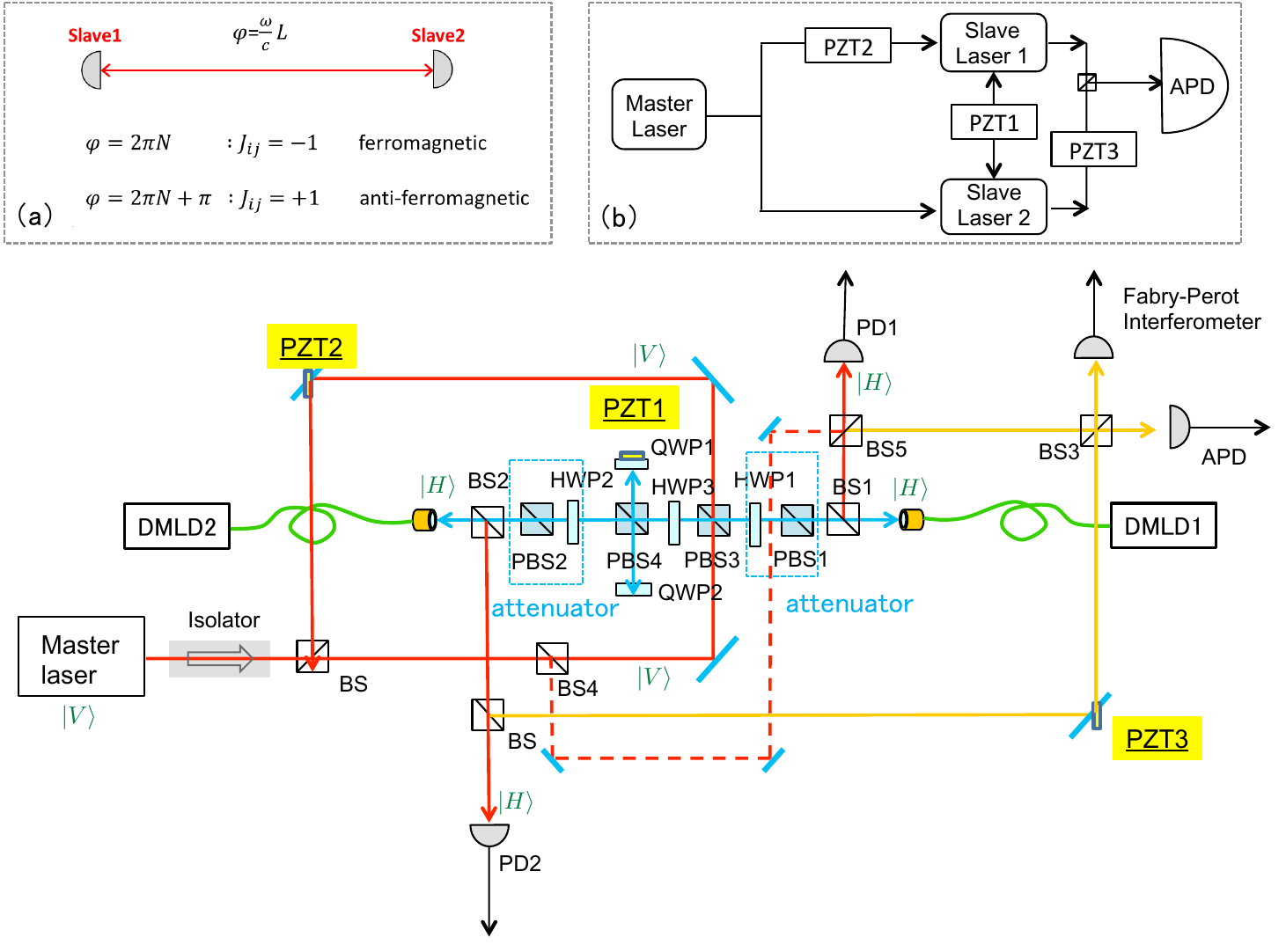}
\caption{Experimental setup. The blue line shows the mutual coupling between two 
slave lasers (DMLDs). The red line shows uni-directional injection from the master 
laser that implements the Zeeman term $\lambda_{iz}$. The relative phase of DMLD1 
and 2 is measured by the Mach-Zehnder interferometer on the yellow lines with the 
FPI and APD. PD1, PD2, APD and FPI each have a refractive neutral density filter 
($\sim$ 10dB attenuators) inserted in front of them as isolators.  [Inset(a)]: How 
to implement the Ising interaction between two slave lasers. In-phase coupling 
between two slave lasers corresponds to ferromagnetic coupling. 
[Inset(b)]:Schematic picture of two site Ising model implementation. }
\label{Fig:2}
\end{figure}
\ \ \ \ %
In our experimental setup, a master laser signal is uni-directionally 
injected into two slave lasers, which also implements the Zeeman term in Eq. 
(\ref{eq:1.1 Ising}) if the phase is slightly rotated towards either 
$+\frac{\pi}{2}$ or $-\frac{\pi}{2}$ from the standard phase zero. 
The slave lasers are mutually coupled by bi-directional injection-locking, 
which implements of the Ising coupling term in Eq. (\ref{eq:1.1 Ising}).
The detailed experimental setup is depicted in Fig. \ref{Fig:2}. The optical path 
length between the two slave lasers and those between each slave and master laser 
are independently controlled by a polarization-dependent optical path. We used 
a tunable external cavity single-mode diode laser as the master laser(Koshin 
Kogaku LS-601A-56S2); it oscillates around 1577.5 nm with a spectral linewidth 
$\leq 30$ kHz. We used discrete mode diode lasers (DMLD, Eblana photonics FLDs) 
with a spectral linewidth  $\leq 100$ kHz as the slave lasers.  

Two slave lasers are mutually coupled along the blue line in Fig. \ref{Fig:2}. 
These slave lasers are nearly identical, and their injection currents are set to 
be the same in the experiment. However, their temperatures must be tuned so as 
to make their self-oscillation frequencies identical. The optical path length 
between the slave lasers is  $L=1550$ mm. Moreover, their optical path length 
is controlled by piezoelectric transducer 1 (PZT1), wherein the modulation 
voltage is used to shift the position of the reflection mirror with a quarter 
wave plate 1 (QWP1). 
The role of the quarter wave plate 1 inserted before PZT1 is to transform the vertical polarization light from the slave laser 1 to the horizontal polarization light before transmitting through PBS4. Likewise, the quarter wave plate 2 transforms the incident horizontal polarization light to the vertical polarization light before reflecting from PBS4. The same transformation is performed onto the output light from the slave laser 2.
The two slave lasers oscillate in the horizontal polarization. 
The polarization beam splitter 1 (PBS1) and half wave plate 1 (HWP1) work together 
as attenuators for the beam from one slave laser (DMLD2), while PBS2 and HWP2 
attenuate the light from the other slave laser (DMLD1). HWP3 is set at 
$\theta=45 ^\circ$  to the polarization axis so that the polarization of the 
incident beam from the slave laser 1 (DMLD1) into HWP3  is rotated to the 
horizontal polarization and it is reflected to QWP1 by PBS4. The master laser 
signal is injected into both slave lasers. The master laser oscillates in the 
vertical polarization and is converted into the horizontal polarization when 
it goes through HWP3. The optical path length for the master laser signal is not 
influenced by PZT1. The phase of the master laser signal injected into slave laser 1 
which comes from the upper branch of the red line is modulated by the mirror 
mounted on PZT2.  The output beams from the slave laser 1 and slave laser 2 are 
partially reflected at BS1 and BS2 and combined at BS3 along the yellow line. 
The PZT3 modulates only optical path length from the slave laser to the BS3 without toughing the optical path length from the slave laser 1 to BS3, so that the interference pattern between the two slave laser outputs can be obtained at APD.
The spectral linewidth and the interferometer output between the two slave lasers 
are detected with a scanning Fabry-Perot interferometer (FPI) and an InGaAs/InP 
avalanche photo diode (APD). As shown in inset(a) of Fig.\ref{Fig:2} , the ferromagnetic 
and the anti-ferromagnetic couplings are implemented by in-phase and out-of phase 
coupling path length between the two slave lasers.The inset(b) of Fig.\ref{Fig:2} 
shows a simplified picture of the experimental setup in Fig.\ref{Fig:2}. A master 
laser signal is injected to the two slave lasers and a path from the master laser 
to the slave laser 1 is modulated by PZT2. The mutual coupling path between the 
two slave lasers is modulated by PZT1. A path difference in the interference between 
the two slave lasers is modulated by PZT3.

\section{Slave laser phase modulation by injection-locking}
\ \ \ \ The phase of the injection-locked slave laser is continuously shifted 
from that of the master laser if there is a frequency detuning between the master 
laser and slave laser within the locking bandwidth \cite{bib: Takata 2012}. 
The phase of an injection-locked slave laser shifts continuously from 
$-\frac{\pi}{2}$ to $\frac{\pi}{2}$ inside the locking bandwidth. Denoting the 
frequency detuning between the slave laser and the master laser by $\Delta \omega$, 
the relative phase shift $\phi_0$ satisfies the following relation:
\begin{eqnarray}
\Delta \omega &\equiv& \omega-\omega_{r0}\\
& = & (\rm sin \phi_0+\alpha \rm cos \phi_0)\frac{F_0}{A_0}\sqrt{\frac{\omega}{Q_e}} ,
\end{eqnarray}\label{eq:4.1 Phase}
where $\omega_{r0} $ is the self-oscillation frequency of the slave laser without 
injection-locking and $\omega$ is the master laser oscillation frequency, 
$F_0$ and $A_0$ are, respectively,  the injection field amplitude from the master 
laser and slave laser internal field amplitude, and $Q_e$ is an external cavity 
quality factor of the slave laser. $\alpha$ is the linewidth enhancement factor. 
$|F_0|^2$ is normalized to express a photon flux, while $|A_0|^2$ is normalized 
to represent a photon number.

\begin{figure}[htbp]
\centering\includegraphics[width=7cm]{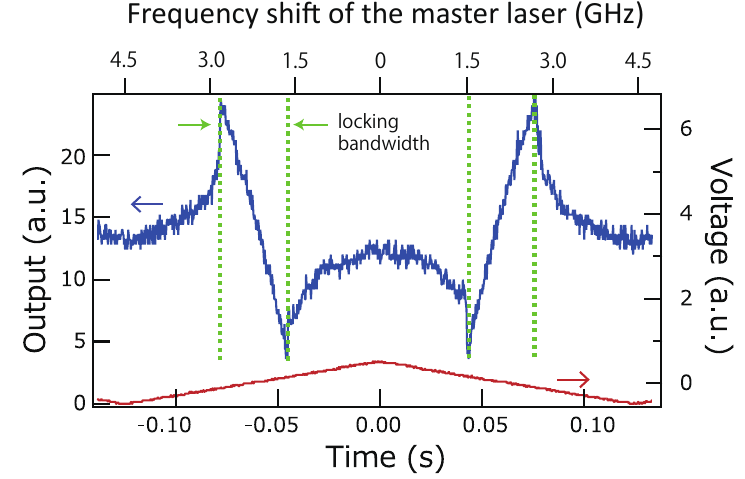}
\caption{The blue trace represents the Interference output between the master laser and slave laser 1 as a function of the frequency shift of the master laser. The red trace is the ramp voltage to sweep the master laser frequency.}\label{Fig:3}
\end{figure}

We measured the phase shift of the slave laser due to the frequency detuning 
$\Delta f=\Delta \omega/2 \pi$ between the master laser and the slave laser within 
the locking bandwidth. The frequency of the master laser is externally modulated 
at a 10-Hz repetition rate through its injection current. In Fig. \ref{Fig:3} 
the red line shows the applied voltage used to modulate the injection current. 
The blue line shows the interference signal between slave laser 1 and the master 
laser observed at BS5  in Fig. \ref{Fig:2} for identical output powers of $1.12$ mW. 
The range of the frequency excursion of the master laser is $4.5$ GHz to cover the 
full locking bandwidth. The observed locking bandwidth is $\sim1.3$ GHz (the 
locking-bandwidth boundaries are indicated by the green dotted lines). Since the 
slave laser has a negligible linewidth enhancement factor, $\alpha \simeq 0$, 
the locking bandwidth is given by 
$\Delta f_{LB} \equiv \frac{1}{2 \pi} \frac{\omega}{Q}\sqrt{\frac{P_{in}}{P_{out}}}$ 
\cite{bib: Kobayashi 1981, bib: Gillner 1990} where $P_{in}=\hbar \omega |F_0|^2$ 
and $P_{out}=\hbar \omega |A_0|^2\frac{\omega}{Q_e}$ are the input and output 
powers of the slave laser. 
However, in the slave laser (DMLD) used in this experiment, most 
of its cavity is formed by a passive waveguide without carriers so that the 
effective linewidth enhancement factor is negligible. This is experimentally 
manifested by the symmetric phase excursion of the slave laser for upward and 
downward frequency detuning shown in Fig. \ref{Fig:3}.
The phase of the slave laser is supposed to shift  from 
$-\frac{\pi}{2}$ to $+\frac{\pi}{2}$ with the frequency detuning between the 
master and slave lasers \cite{bib: Gillner 1990}. 
The relative phase shift shown in Fig. \ref{Fig:3} is monotonical with the frequency 
detuning, which indicates the linewidth enhancement factor $\alpha$ of the slave 
laser (DMLD) is indeed small. 
\\

\ \ 
\section{Numerical simulation for two mutually coupled slave lasers}
\ \ \ \ In the phase model of an injection-locked laser network, the binary phase 
of each slave laser acts as an Ising spin \cite{bib: Utsunomiya 2011}. The c-number 
Langevin equations of the time-dependent amplitude $A_{i}(t)$, phase $\phi_{i}(t)$, 
and carrier number $N_{Ci}(t)$ for an i-th slave laser are described as,
\begin{eqnarray}
\frac{d}{dt}A_{i}(t) &=& -\frac{1}{2} \Big[\frac{\omega}{Q}-E_{CV_i}(t) \Big]A_{i}(t)+\frac{\omega}{Q}\sqrt{n_M}\{\zeta \cos \phi_{Vi} (t)-\eta \lambda_i \sin \phi_{Vi}\}  \nonumber \\
&& -\frac{\omega}{Q}\sum_{j\neq i}\frac{1}{2}\eta J_{ij}A_{j}(t)\cos (\phi_{j}(t)-\phi_{i}(t))+F_{Ai} ,\\
\frac{d}{dt}\phi_{i}(t) & = & \frac{1}{A_{i}(t)}\Biggl\{ \frac{\omega}{Q} \sqrt{n_M}[-\zeta \sin \phi_{Vi}(t)-\eta \lambda_i \cos \phi_{Vi}(t)]  \nonumber\\
&& -  \frac{\omega}{Q} \sum_{j \neq i} \frac{1}{2} \eta J_{ij} A_{j}(t)\sin(\phi_{j}(t)-\phi_{i}(t))  \Biggr\} +F_{\phi i} ,\\
\frac{d}{dt}N_{i}(t)&=&P-\frac{N_{i}(t)}{\tau_{sp}}\biggl\{1+\beta \big[A_{i}(t)^2+1\big]\biggr\}+F_{Ni} .
\end{eqnarray}\label{eq:4.2 Numerical Simulation}
Here $F_{Ai}$, $F_{\phi_i}$ and $F_{Ni}$ are the Langevin noise terms for the 
amplitude, phase and carrier number. $\omega / Q$ is the cavity photon decay rate, 
$\zeta$ is the optical attenuation coefficient of the master laser injection signal, 
$\eta$ is the common attenuation coefficient of the Zeeman term and Ising term, 
$\tau_{sp}$ is the spontaneous emission lifetime, $P$ is the pump rate into the 
slave laser i , $E_{CVi}=\beta N_i /\tau_{sp}$ is the spontaneous emission rate 
for i-th slave laser. $\beta$ is the fractional coupling efficiency of spontaneous 
emission into a lasing mode \cite{bib: Utsunomiya 2011}. 

\begin{figure}[tbp]
\centering\includegraphics[width=13cm]{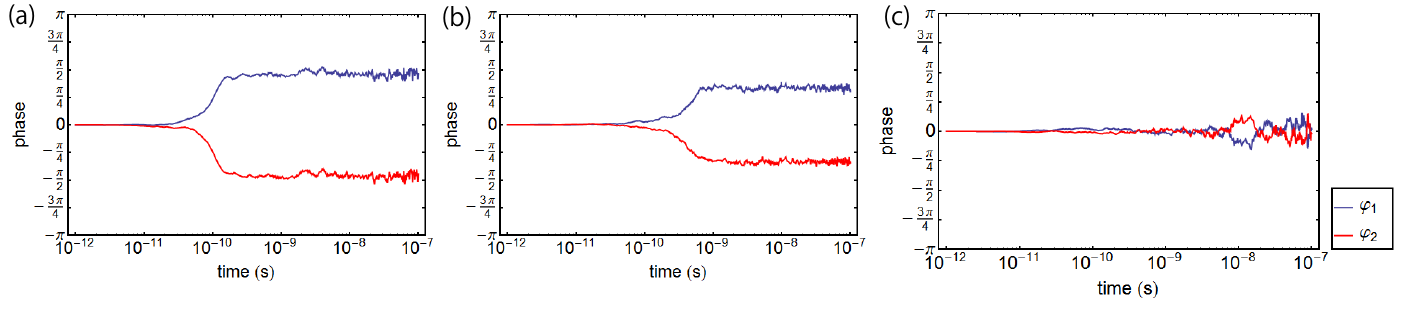}
\caption{Bifurcation of two slave lasers' phases in mutually coupled semiconductor lasers: (a)The Ising term is dominant when the mutual coupling between the slave lasers is relatively strong ($\eta=0.04$) 
(b)The mutual coupling ratio for the Ising term is weaker than the case of (a) ($\eta =0.01$). 
(c)The master injection term is dominant when the mutual coupling is relatively small ($\eta =0.005$). 
The other numerical parameters are $\lambda_1=\lambda_2=0$, $\omega / Q=10^{12}$( s$^{-1}$), $\tau_{sp}=10^{-9}$(s), $\zeta=0.005$ and $\beta=10^{-4} $}\label{Fig:4}
\end{figure}

Figure \ref{Fig:4} shows numerical simulation results for the time-dependent phases 
of the two slave lasers, where the Ising coupling $J_{12}$ is $ +1$ (anti-ferromagnetic 
coupling).  In this case, the steady state solution is an anti-ferromagnetic order, which 
means the two slave lasers have opposite phases. In the numerical simulation, 
the slave laser phases are initially set to the reference phase (zero), which is determined 
by the master laser injection signal. When the mutual coupling term is switched on, 
as shown in Fig. \ref{Fig:4}(a), the slave laser phases bifurcate into $+\frac{\pi}{2}$ and $-\frac{\pi}{2}$. 
On the other hand, when the injection from the master laser is relatively strong 
compared to the mutual coupling, as shown in Fig. \ref{Fig:4}(c), the slave laser 
phases stay in the reference phase (zero). When the mutual coupling between two slave lasers is slightly decreased, the relative phase difference between them is smaller than $\pi$ as shown in Fig. \ref{Fig:4}(b).
It is assumed that the frequency of the two slave lasers are locked to the master 
laser frequency $\omega$ in these numerical simulations.

\ \
\section{Switching between ferromagnetic and anti-ferromagnetic phase orders}
\begin{figure}[htbp]
\centering\includegraphics[width=6.5cm,bb=0 0 403 214]{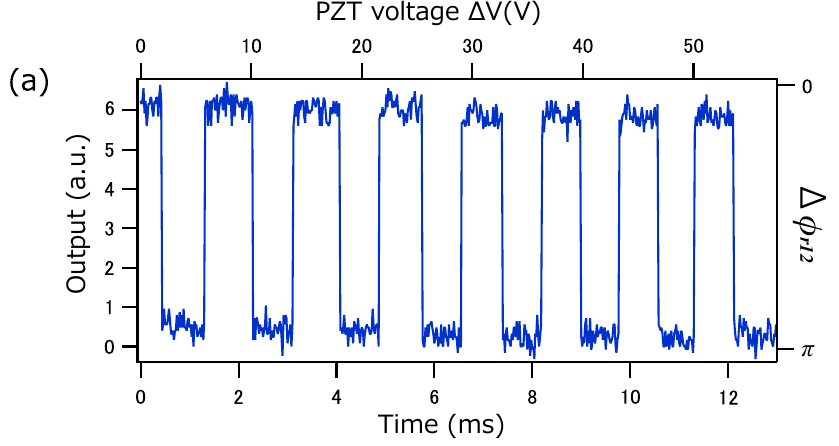}
\centering\includegraphics[width=6cm,bb=0 0 394 208]{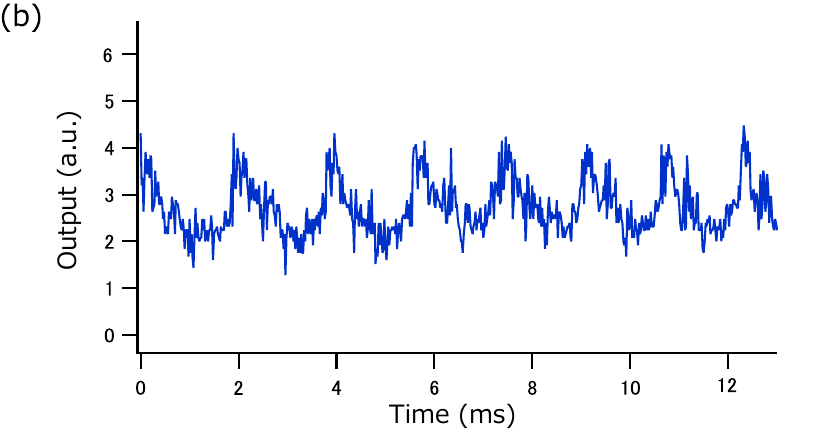}
\centering\includegraphics[width=6cm,bb=0 0 394 208]{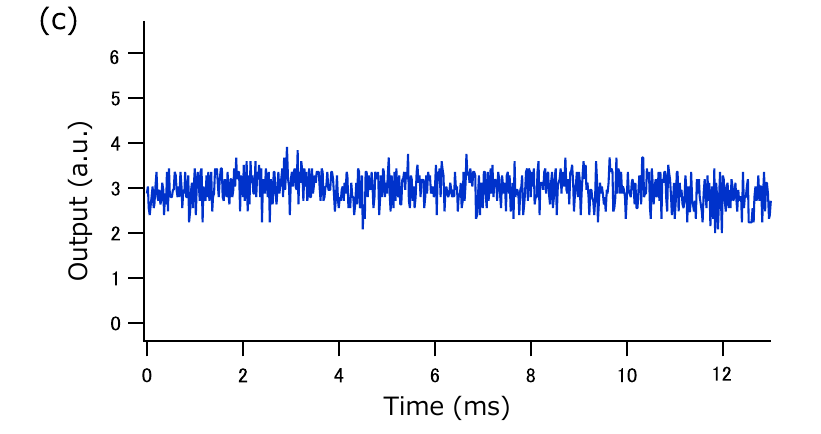}
\caption{Interference signal output to show the relative phase $\Delta \phi_{r12}$ between two slave lasers (DMLD1 and 2) when the mutually coupling optical path length is modulated: (a) Ising coupling term is dominant, (c) Master injection term is dominant, (b) Transition regime between (a) and (c).}\label{Fig:5}
\end{figure}

We experimentally studied how the slave laser phases are modulated when the optical 
path length between the two slave lasers (DMLDs) is varied. Notice that the optical 
path length modulates the polarity of the Ising coupling term $J_{ij}$. 
We observed a transition from one regime, where the Ising coupling is dominant, 
to the other regime, where the master injection term is dominant, 
when we changed the ratio of the uni-directional injection from the master laser 
into the two slave lasers to the mutual coupling between the two slave lasers by rotating HWP1 or HWP2 
in Fig. \ref{Fig:2}. Figure \ref{Fig:5} (a) shows the interference signal between 
the two slave lasers (DMLD1 and 2) when there is relatively strong mutual coupling 
between them, with $P_{in M_1}=1.5 \mu W$ and $P_{in S_{12}}=33 \mu $W. Here 
$P_{in M_1}$ is the injection power from the master laser into a fiber coupler of 
the slave laser 1 and $P_{in S_{12}}$ is the injection power from the slave laser 2 
into the fiber coupler of the slave laser 1. 
It is observed that the strongly coupled slave lasers prefer to oscillate in 
discrete phases. The relative phase difference between two slave lasers seems to 
be either 0 or $\pi$, as demonstrated by an interference signal in Fig. \ref{Fig:5}(a). 
The frequencies of the two slave 
lasers now deviate from that of the master laser, as described in the next section. 
We observed that the slave lasers still oscillate at identical frequencies, 
which is true even after the injection master signal is turned off. 
The voltage applied to PZT1 linearly increases as a function of time in Fig. \ref{Fig:5}. 
The PZT1 displacement is $\Delta d$ / $\Delta V$ =12.23$ \pm 2.5 \times 10^{-8}$ m/V 
with respect to the applied voltage $\Delta V$. 
The optical path modulation for mutual coupling causes the phase difference between 
the two slave lasers to have discrete jumps every $\Delta V \sim 3.6$V. 
The high and low outputs correspond to the ferromagnetic and anti-ferromagnetic 
phase orders. 
When the master laser injection and the mutual coupling signals are set to be 
$P_{in M_1}=2.2 \mu W$ and $P_{in S_{12}}=0.3 \mu W$, the interference 
signal doesn't show any modulation against the mutual coupling path modulation (Fig. \ref{Fig:5}(c)).  
When the two signal intensities, are set to be on the same order ($P_{in S_{12}}=2.0 \mu $W) as shown in Fig. \ref{Fig:5}(b), the interference signal shows a transition between the mutual coupling dominant regime in Fig. \ref{Fig:5}(a) and the master injection dominant regime in Fig. \ref{Fig:5}(c).

\begin{figure}[tbp]
\centering\includegraphics[width=6cm]{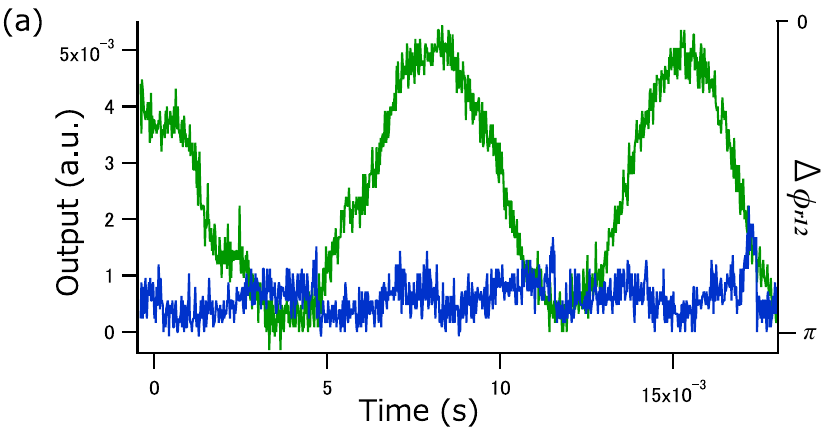}
\centering\includegraphics[width=6cm]{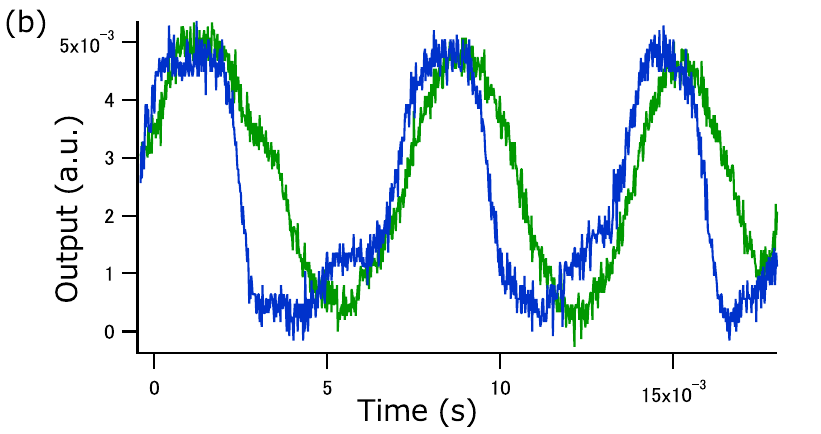}
\caption{Interference signals to show the relative phase $\Delta \phi_{r12}$ 
between the two slave lasers (DMLD1 and DMLD2). The green lines show the case 
that the interference path length is modulated. The blue lines show the case that 
the injection phase from the master laser to the slave laser 1 is modulated.  
(a) The Ising coupling term is dominant under a weak master signal injection. 
(b) The master signal is dominant under a strong master signal injection. }
\label{Fig:6}
\end{figure}

Figure \ref{Fig:6} shows the interference output when the two slave laser signals 
are combined at BS3 and detected by the APD as shown in Fig. \ref{Fig:2}.  
The blue line shows the interference signal when PZT2 is electrically driven 
linearly as a function of time, which causes the phase of the master laser 
signal injected into the slave laser 1 (DMLD1) to be modulated. The blue trace in 
Fig. \ref{Fig:6}(a) shows the case of using the same experimental parameters 
of  $P_{in M_1}=1.5 \mu$W and $P_{in S_{12}}=33 \mu $W as Fig. \ref{Fig:5}(a), 
where the mutual coupling is stronger than the uni-directional master signal 
injection. Here, the phase modulation of the injected master laser signal is not 
effective. The green trace in Fig.\ref{Fig:6}(a) shows the interference signal when 
only PZT3 is linearly modulated as a function of time, which manifests the phase 
coherence between the two slave lasers. The blue trace in Fig. \ref{Fig:6}(b) 
shows the interference signal with PZT2 modulation under a relatively strong master 
signal injection ($P_{in M_1}=147 \mu$W) compared with the mutual injection power 
($P_{in S_{12}}=33 \mu $W). Here, the interference signal has the same phase 
modulation of the master laser. It suggests that the oscillating phase of slave 
lasers are mainly determined by the injected master laser signal and 
there's phase coherence between two slave lasers as 
shown by the green trace in Fig. \ref{Fig:6}(b). 
\\
\\
\section{Spontaneous frequency selection of mutually coupled slave lasers}
\begin{figure}[htbp]
\centering\includegraphics[width=10cm,bb=0 0 592 208]{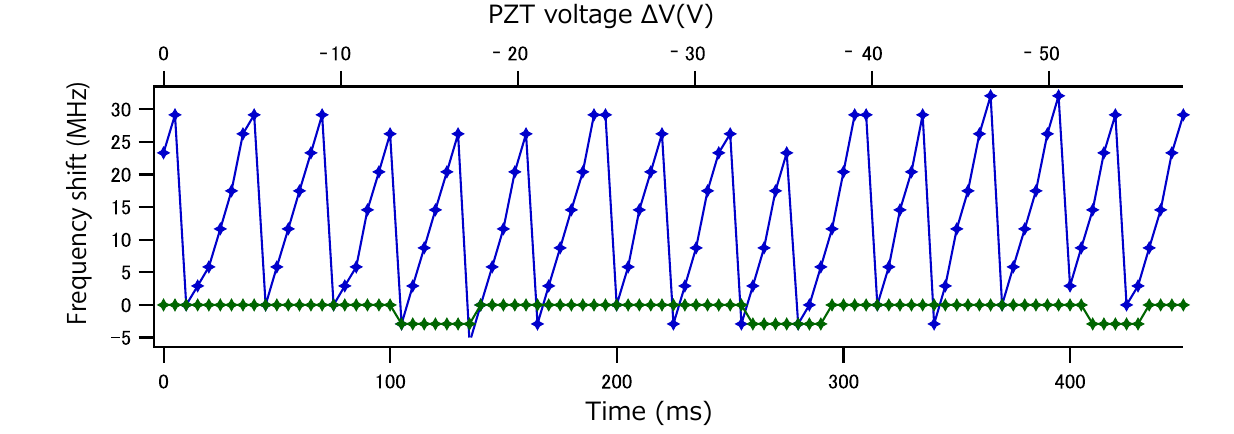}
\caption{Frequency shift of a slave laser due to the mutual coupling path length 
modulation as a function of time under a strong mutual injection (blue line) and 
a weak master signal injection (green line).}\label{Fig:7}
\end{figure}
We observed a sawtooth variation in the slave laser frequency when the optical path 
length between the slave lasers is modulated, as shown by blue dots in Fig. \ref{Fig:7}. 
Here the mutual coupling is dominant over the master laser injection.
The experimental results shown in Fig. \ref{Fig:6}(a) indicate that the oscillation 
frequencies of the two slave lasers are synchronized in the strong mutual coupling 
regime. We measured the frequency shift of the slave laser 1 (DMLD1) with the 
FPI at the top right corner in Fig. \ref{Fig:2} , while the signal from the slave 
laser 2 (DMLD2) is blocked just before BS3. Figure \ref{Fig:7} plots the slave 
laser frequency shift against the mutual coupling path length modulation between 
the two slave lasers. When the mutual coupling is stronger than the master laser 
signal injection, a sawtooth frequency shift of $ \sim 30$ MHz is observed as 
shown by the blue line in Fig. {\ref{Fig:7}}. In this case, the phases are pinned 
either in the same phase or in the opposite phase, as shown in Fig. \ref{Fig:5}(a). 
On the other hand, the oscillation frequency of the slave laser is held constant, 
when the master laser injection is dominant, as shown by the green line in 
Fig. {\ref{Fig:7}}. Here discrete jumps in the oscillation frequency are experimental 
artifacts due to the discretization error(resolution limit) of the FPI. 
There are $N=21$ sawtooth-frequency-shift cycles when the applied voltage is swept 
over $\Delta V=78$V, so that frequency shift occurs at every applied voltage of 
$\Delta V= 3.7$V, which corresponds to a PZT1 shift of $\Delta d_L=456.3 \pm 92.8$nm. 
This sawtooth-frequency-shift is considered to be the result of spontaneous 
selection of the synchronized oscillation frequencies of the two slave lasers to 
minimize the overall threshold pump current or to maximize the output power  
\cite{bib: Saito 1982}. 
The favorable phase difference between the two slave lasers is either 
$\theta_1 - \theta_2 =0$ or $\pi$ so that the two slave lasers build up a standing 
wave between the edges of the two slave lasers. 
\begin{figure}[tbp]
\centering\includegraphics[width=5.5cm,bb=0 0 467 230]{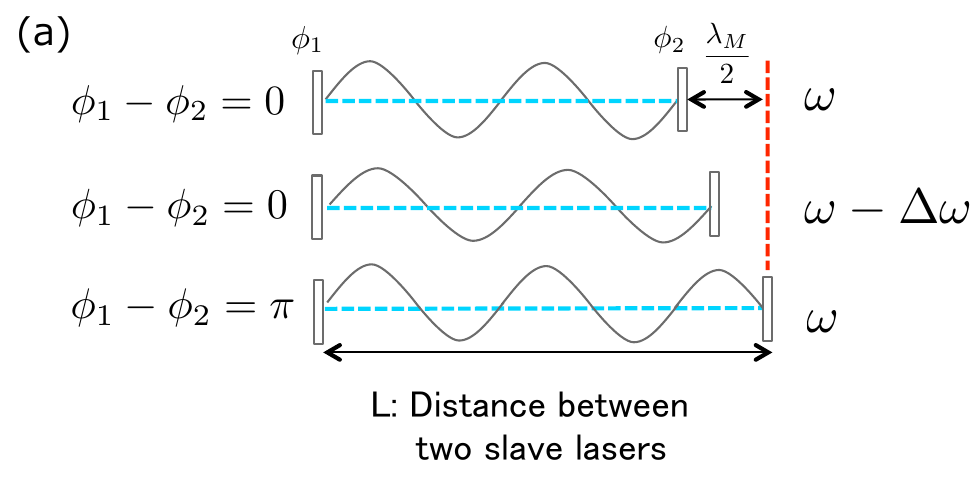}\ \ \ \ \ 
\centering\includegraphics[width=6.5cm,bb=0 0 388 195]{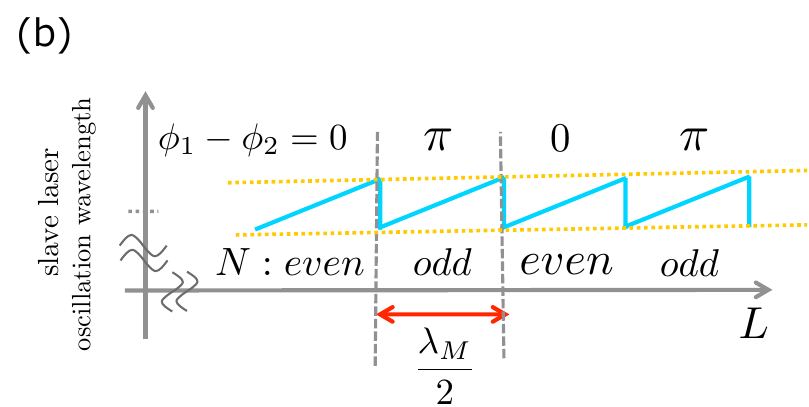}
\caption{Schematic explanation for the sawtooth frequency modulation and phase jump in two coupled slave lasers, which explain the experimental results shown in Fig. \ref{Fig:5}a and Fig. \ref{Fig:7} simultaneously.}\label{Fig:8}
\end{figure}

Figure {\ref{Fig:8}} is a schematic explanation of the spontaneous frequency 
optimization. When the optical path length between the two slave lasers is an 
integer multiple of the self-oscillation wavelength $\lambda_M=2\pi c/\omega$, 
the number of loops of the standing wave is even, so that the two slave lasers 
oscillate with the same phase. In the second figure of Fig. {\ref{Fig:8}}(a), 
the oscillation wavelength become slightly elongated as the optical path length 
$L$ is extended while the phase difference between the two slave lasers is still 
$\phi_1-\phi_2=0$. However, when $L$ is shifted $\frac{\lambda_M}{2}$ from the 
resonant condition in the first figure of Fig. {\ref{Fig:8}}(a), the oscillation 
frequency returns to the self-oscillation frequency $\omega$  and the number of 
the loops of the standing wave between the two facets is odd, so that the phase 
difference between the two slave lasers is $\phi_1-\phi_2=\pi$. We plot that 
description of oscillation frequency shift as a function of $L$ in Fig. {\ref{Fig:8}}(b). 
$N$ is the number of the loops and corresponding phase differences $\phi_1-\phi_2$, 
are written at the top. The light blue line shows the slave laser oscillation 
frequency shift according to the linear shift of $L$. This frequency shift for 
a path length variation $\Delta L=\frac{\lambda_M}{2}$ corresponds to the free 
spectral range $\Delta \nu_{FSR}=\frac{c}{2L}$ at the optical cavity length $L$. 

The path length variation for a single frequency shift event in our experiment 
as shown in Fig. \ref{Fig:5} is $\Delta L=\frac{\lambda_M}{4}$ because the path 
length between the two slave lasers changes twice when the position of the mirror 
just before QWP is changed. The expected position shift of PZT1 for a single 
sawtooth frequency shift is $\Delta L=\frac{\lambda_M}{4}=394.4$nm which is close 
to the measured position shift of PZT1 $\Delta d_L=456.3 \pm 92.8$nm. The optical 
path length between the two slave lasers in Fig. \ref{Fig:2} is $L=1550$mm. The 
sawtooth modulation in Fig. {\ref{Fig:7}} is $\sim 30$ MHz, which is less than 
30 percent of the estimated frequency shift $\Delta \nu_{FSR}=\frac{c}{2L}=96.8$MHz. 
The difference between the experimentally observed frequency excursion and the free 
spectral range stems from the phase shifts associated with the reflection from the 
two slave laser facets. As indicated in Eq. (3), the slave laser phase shifts away 
from the injection signal phase when there is a detuning between the self-oscillation frequency $\omega_{r0}$ and the actual oscillation frequency $\omega$.

\section{Conclusion}
\begin{figure}[htbp]
\centering\includegraphics[width=10cm]{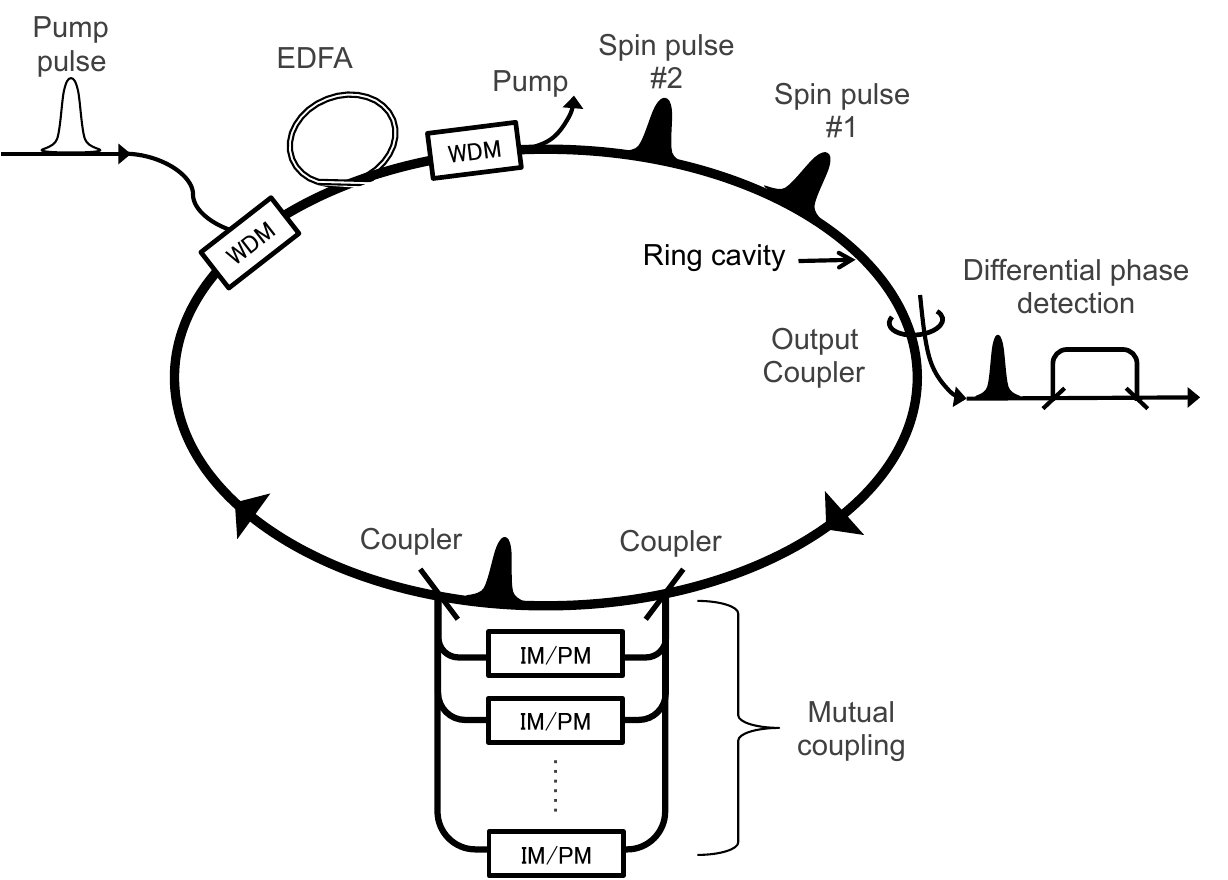}\ \ \ 
\caption{Ising machine based on a higher harmonic fiber mode-locked laser with $N-1$ delay lines and the deferential phase detection.}\label{Fig:9}
\end{figure}
\ \ We have implemented a two-site Ising model by using the two mutually coupled 
slave lasers that are simultaneously injection-locked by a single master laser. 
As shown in Fig. \ref{Fig:5}(a), the two slave lasers feature the ferromagnetic and anti-ferromagnetic orders when the two lasers are coupled in-phase ($J_{ij} < 0$) or out-of-phase ($J_{ij} >0$), which is the ground state of the Ising Hamiltonian. 
The transition from the Ising-term dominant regime to the master-injection-term 
dominant regime was observed. The phase of the slave laser is continuously 
synchronized  with that of a master laser when the master laser injection is 
stronger than the mutual coupling between the slave lasers.  On the other hand, 
when the mutual coupling is stronger than the master laser injection, the phases 
of the two slave lasers jump discretely and the oscillation frequencies continuously 
shift with the optical path length modulation between the two slave lasers. In 
that case, the mutually coupled slave lasers communicate with each other and 
oscillate at the optimum frequency, where a standing wave develops between them 
that results in either a ferromagnetic or anti-ferromagnetic phase order. We note 
that as related work, the dynamic and chaotic oscillation behaviors of mutually 
coupled semiconductor lasers have been studied numerically and experimentally 
\cite{bib: Rogister 2004}\cite{bib: Fujino 2001}. 

€œFor practical implementation of a larger-scale laser network based Ising machine, we can use a higher harmonic fiber mode-locked laser as shown in Fig. {\ref{Fig:9}}. In this implementation, $N$ optical pulses are simultaneously produced in a single fiber ring cavity so that they have identical oscillation frequencies. In order to implement the Ising coupling term, $J_{ij} \sigma_{iz} \sigma_{iz}$, we can introduce $N-1$ optical delay lines with independent intensity and phase modulators. The pick-off signal from i-th pulse is properly modulated in its intensity and phase, and coupled to the j-th pulse with an appropriate delay time. This time division multiplexing scheme allows us to implement $N(N-1)/2$ mutual coupling paths by only $N-1$ delay lines. Finally, the computational results are recorded by the one-bit delay differential phase detection \cite{bib: Marandi 2014}.
When implementing with a higher harmonic mode-locked fiber laser with 1GHz repetition frequency and 20m fiber ring cavity, 
the total number of optical pulses is $\sim$100 in this system. 
If we increase the clock frequency up to 10GHz and the fiber length up to 200m, we can implement $N\sim$ 10,000 optical pulses.
\\
\section{Acknowledgments}
\ \ The authors would like to thank S. Tamate for fruitful discussions. This research is supported by the Cabinet Office, Government of Japan and the Japan Society for the Promotion of Science (JSPS) through the Funding Program for World-Leading Innovative R\&D on Science and Technology (FIRST Program) and a Matsuo academic grant and Shiseido Female Researcher Science Grant.\\ \\
\ \ This paper was published in Optics Express and is made available as an electronic reprint with the permission of OSA. The paper can be found at the following URL on the OSA website: http://www.opticsinfobase.org/oe/abstract.cfm?uri=oe-23-5-6029 . Systematic or multiple reproduction or distribution to multiple locations via electronic or other means is prohibited and is subject to penalties under law.
\end{document}